\documentclass[12pt,preprint]{aastex}
\usepackage{epsfig}
\usepackage{psfig}

\newcommand{\chisq}{$\chi^2$}

\newcommand{\PSRB}{PSR B0628$-$28}
\newcommand{\RX}{RX J0630.8$-$2834}
\newcommand{\CXC}{CXC J063049.4$-$283443}

\shorttitle{{A Multi-wavelength study of PSR B0628-28}}
\shortauthors{W.~Becker et al.}

\begin{document}

\title{{A Multi-wavelength study of PSR B0628-28:\\[1ex] The first over-luminous rotation-powered pulsar?}}

\author{
Werner Becker\altaffilmark{1},
Axel Jessner\altaffilmark{2},
Michael Kramer\altaffilmark{3},
Vincenzo Testa\altaffilmark{4},
Clemens Howaldt\altaffilmark{1}}   

\altaffiltext{1}
{Max-Planck Institut f\"ur Extraterrestrische Physik, 85741 Garching bei M\"unchen, Germany}
\altaffiltext{2}
{Max-Planck Institut f\"ur Radioastronomie, Effelsberg, 53902 Bad M\"unstereifel, Germany}
\altaffiltext{3}
{University of Manchester, Jodrell Bank Observatory, Macclesfield, Cheshire SK11 9DL, UK}
\altaffiltext{4}
{INAF - Osservatorio Astronomico di Roma, Via Frascati 33  00040 Monte Porzio Catone (Italy)}

\begin{abstract}
 The ROSAT source \RX\, was suggested by positional coincidence to be
 the X-ray counterpart of the pulsar \PSRB. This association, however,
 was regarded to be unlikely based on the computed energetics of the
 putative X-ray counterpart. In this paper we report on
 multi-wavelength observations of \PSRB\, made with the ESO/NTT
 observatory in La Silla, the Lovell telescope at Jodrell Bank and
 XMM-Newton.  Although the optical observations do not detect any
 counterpart of \RX\, down to a limiting magnitude of V=26.1 mag and
 B=26.3 mag, XMM-Newton observations finally confirm it to be the
 pulsar's X-ray counterpart by detecting X-ray pulses with the radio
 pulsar's spin-period. The X-ray pulse profile is not sinusoidal but
 characterized by a two component pulse profile, consisting of a broad
 peak with a second narrow pulse leading the main pulse by $\sim
 144^\circ$. The fraction of pulsed photons is $(39 \pm 6)\%$ with no
 strong energy dependence in the XMM-Newton bandpass. The pulsar's
 X-ray spectrum is well described by a power law with photon index 
 $\alpha=2.63^{+0.23}_{-0.15}$. A composite Planckian plus 
 power law spectral model yields an interesting alternative which 
 formally describes the observed energy spectrum equally well. Inferred
 from best fits are a blackbody temperature of  $\sim 1.7\times 10^6$ K
 and a projected blackbody radius of $\sim 69_{-25}^{+30}$\,m, yielding
 a thermal flux contribution of $\sim 20\%$ within the $0.1-2.4$ keV band. 
 The pulsar's spin-down to X-ray energy conversion efficiency as obtained 
 from the single power law spectral model is $\sim 16\%$, aumming the distance 
 inferred from the radio dispersion measure. If confirmed, \PSRB\, would be the 
 first X-ray over-luminous rotation-powered pulsar identified among all $\sim 1400$ 
 radio pulsars known  today. The emission beam geometry of \PSRB\, is estimated from 
 radio polarization data taken at 408 MHz and 1400 MHz. A formal best fit of the 
 1400 MHz data yields $\alpha\sim 11^\circ$ for the inclination of the magnetic axis 
 to the rotation axis and $\beta\sim -3^\circ$ for the impact angle of the 
 line of sight. A combination of results obtained from 408 MHz and 1400 MHz 
 data, however, makes a nearly orthogonal solution with $\alpha\sim 70^\circ$ and 
 $\beta\sim -12^\circ$ the most likely one. 

\end{abstract}


\section{INTRODUCTION \label{intro}}

 \PSRB\, is a bright radio pulsar (Large, Vaughan \& Wielebinski,
1969). Its characteristic spin-down age is $\sim 2.75 \times 10^6$
years. The pulsar's period and period derivative are $P=1.24$ s and
$\dot{P}=7.1\times 10^{-15}$, implying a spin-down luminosity of $\log
\dot{E} = 32.16\,\,\mbox{erg s}^{-1}$ and a magnetic field at the
neutron star poles of\, $\log B_\perp=12.48$ G. The distance to the
pulsar as listed in the Princeton Pulsar Catalog (Taylor, Manchester
\& Lyne 1993) and the ATNF online pulsar database is 2.14 kpc (Taylor
\& Cordes 1993; Manchester et al.~2005). Using the NE2001 Galactic
free electron density model of Cordes \& Lazio (2002) which
builds upon and supersedes the Taylor \& Cordes (1993) model by
exploiting new observations and methods, the radio dispersion measure
of $\mbox{DM}=34.36\, \mbox{pc cm}^{-3}$ implies that the pulsar might
be closer at a distance of only 1.45 kpc.
 The pulsar parameters and the emission properties observed in other
 X-ray detected rotation-powered pulsars (cf.~Becker \& Tr\"umper
 1997; Becker \& Aschenbach 2002 and references therein), seem to
 indicate that \PSRB\ is an ordinary, old\footnote{in standards of
 X-ray detected non-recycled pulsars} radio pulsar which is
 detectable just at the threshold of sensitivity of the current
 generation of X-ray observatories.  The results from ROSAT and
 Chandra observations, however, suggest that there might be more to
 expect from this pulsar.

 \PSRB\, was first noticed to have a faint potential X-ray counterpart
 in the ROSAT all-sky survey, \RX\, (Becker et al.~1993). The pulsar
 was therefore re-observed in October 1992 and April 1993 in two short
 pointed ROSAT observations of $\sim 2.5$ ksec and $\sim 6.3$ ksec
 exposures. Both observations confirmed the detection of the X-ray
 source \RX\, and collected 27 source counts from it in the ROSAT
 PSPC (Position Sensitive Proportional Counter). The low number of 
 counts did not allow to obtain more than a rough hardness 
 ratio\footnote{Ratio of the number of counts detected in the soft 
 spectral bands vs.~that detected in the hard bands.} which suggested 
 that \RX\, has a medium soft X-ray spectrum. A timing analysis was 
 not possible either, thus leaving the association between \RX\, and 
 \PSRB\, an open issue and based on a rough positional agreement only.

 Such an association, however, was regarded to be unlikely based on
 the estimated energy budget of the putative X-ray counterpart. The
 X-ray luminosity of \RX\, extrapolated to the pulsar's distance was
 found to be $\sim 37\%$ of the pulsar spin-down energy. Its X-ray
 conversion factor thus would be few hundred times higher than the
 average X-ray conversion factor $\sim 10^{-3}\,\dot{E}$ which is
 observed in the $0.1-2.4$ keV band from other X-ray detected
 rotation-powered pulsars (Becker \& Tr\"umper 1997).

 Follow-up optical imaging and spectroscopic observations with the ESO
 New Technology Telescope (NTT) in La Silla (Chile) were performed to 
 assess the nature of the unidentified X-ray source \RX. Various optical 
 sources were identified in the $\sim 25"$ wide ROSAT PSPC error box, 
 which could not be conclusively ruled out as potential optical counterparts 
 of \RX.

 Recently, \"Ogelman and Tepedelenlioglu (2002) have observed \RX\, as
 part of the Chandra-AO3 guest observer program. The source, \CXC, was
 clearly detected with the ACIS$-$S\footnote{See 
 \protect{$\mbox{http:}//\mbox{cxc.harvard.edu}/\mbox{cdo}/\mbox{user\_guide.html}$} 
 for a detailed description of Chandra, its instrumentation and various
 observing modes.} detector in an 17 ksec exposure in
 which 189 source counts were recorded from it.  Although representing
 an improvement over the ROSAT observations, the low number of source
 counts did not support a detailed spectral analysis which could
 discriminate between various possible spectral models. Fixing the
 column absorption to a value corresponding to a neutral hydrogen
 density of $1.07\times 10^{21}\,\mbox{cm}^{-2}$ (as deduced from the
 pulsar's radio dispersion measure by assuming 10 H-Atoms per free
 electron), both a power law model with photon-index $\alpha=2.45\pm
 0.15$ and a thermal blackbody model with $T=(4.53\pm 0.13) \times
 10^6$ K and $R_{bb}\sim 1\,\mbox{km}$ describe the energy spectrum of
 \CXC\, equally well.  As the frame time of the ACIS-S in the selected
 mode was close to the pulsars rotation period, a timing analysis did
 not yield a significant result and prevented the derivation of
 reliable upper limits for the fraction of pulsed photons. However,
 the high spatial resolution of Chandra along with its superior
 pointing accuracy allowed the determination of a much more accurate
 position of the putative X-ray counterpart \RX\ than before.  The
 Chandra position of RX J0630.8-2834 was found to match the position
 of \PSRB\, by 1.5 arcsec! Any offset from the pulsar radio position
 due to its proper motion (Brisken et al.~2003) is well within the
 uncertainties of the Chandra position. Hence, the positional
 coincidence between \RX\, (\CXC\, resp.) strongly suggests that this
 X-ray source is indeed the counterpart of \PSRB.

 In this paper we report on X-ray, optical and radio observations of
 \PSRB\, which were made with XMM-Newton, the ESO NTT in La Silla and
 the Lovell telescope at the Jodrell Bank Observatory in order to
 explore the spectral and timing emission properties of this
 interesting pulsar. The paper is organized in the following manner:
 in \S2 we describe the optical, radio and XMM-Newton observations of
 \PSRB\, and provide the details of the data processing and data
 filtering. The results of the spectral and timing analysis are given
 in \S3. A summary and concluding discussion is presented in \S4.

\section{OBSERVATIONS AND DATA REDUCTION\label{obs}}

\subsection{OPTICAL OBSERVATIONS OF \PSRB\label{optical_obs}}

 Two images of 600s exposure
 time in the B and V filters, pointed on the position of \PSRB, were
 made on Dec.~15, 1999 with the ESO NTT Multi-Mode Instrument (EMMI). 
 The $B$ (422.3 nm) image was taken with the blue arm,
 equipped with a Tektronix $1024 \times 1024$ chip having 24 $\mu$
 pixels size (0.37 arcsec), while the $V$ (542.6 nm) image was
 acquired with the red arm mounting a Tektronix $2086 \times 2048$
 chip with 24 $\mu$ pixel size (0.27 arcsec).  Two images of the
 Landolt field Ru149 were observed soon after for the purpose of
 calibrating the two images.

 Bias and flat-field were applied as usual, using calibration
 sets acquired at evening twilight, and the reduction of the images
 was performed with both DAOPHOT (Stetson 1987) and SExtractor
 (Bertin and Arnouts, 1996) to take into account the presence of
 extended sources and to check the output photometry
 results. Photometric calibration was performed by using the
 Ru149 field and the average La Silla extinction terms (-0.22 for B
 and -0.12 for V, as found in the instrument web-page at ESO La Silla
 site). The night was photometric and with a good seeing, 0.75 arcsec
 in V.

\begin{figure}
\centerline{\psfig{figure=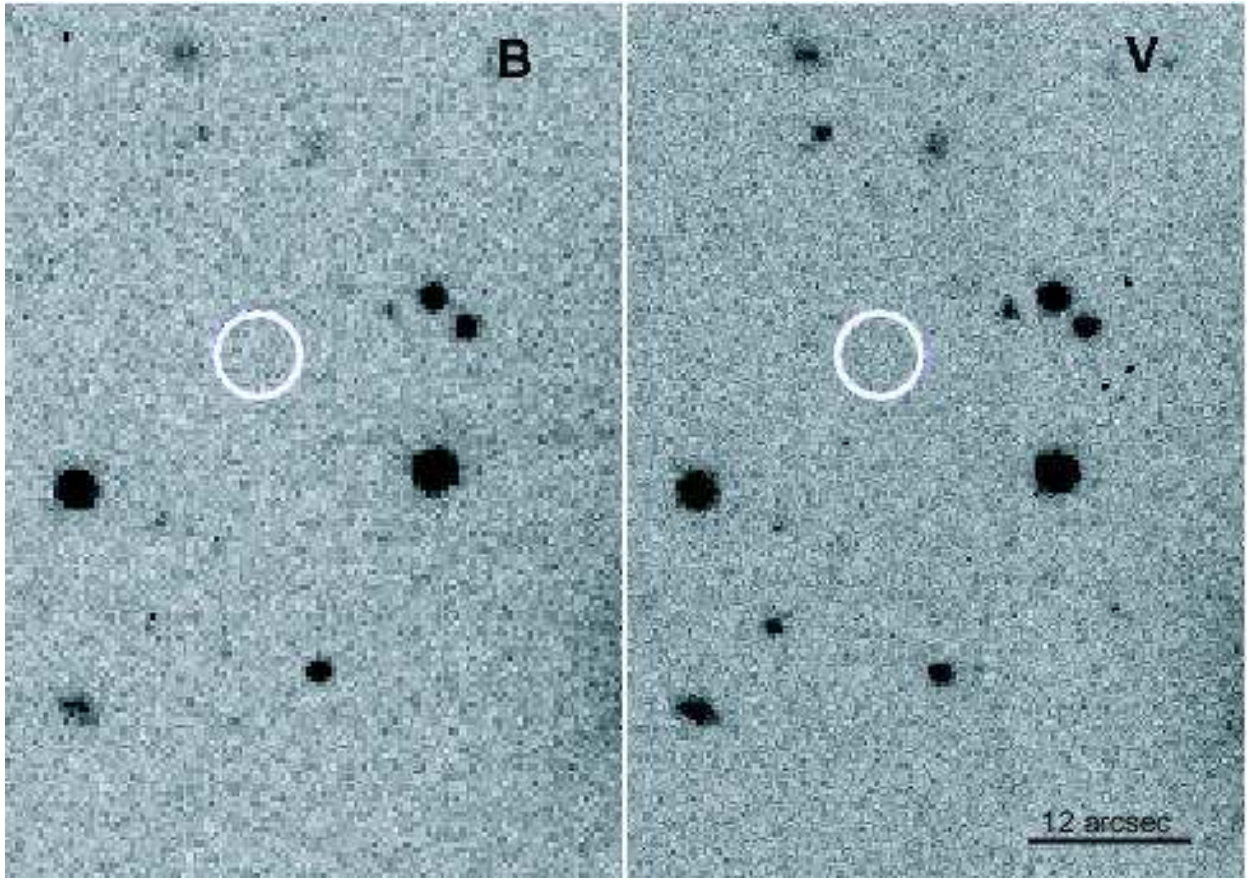,width=16.5cm,clip=}}
\caption[]{The visible sky around \PSRB\, as seen with the ESO NTT/EMMI
telescope/detector in La Silla using the V (542.6 nm) and B (422.3 nm) band
filters. The circles have a radius of 3 arcsec and are centered at the pulsar
position. Top is north and left is east.} \label{optical_images}
\end{figure}

 No source is detected at the position of \PSRB. Figure
 \ref{optical_images} shows a close-up view of the pulsar field as
 seen in the visible (V) and blue (B) bands. 3$\sigma$ upper limits
 assuming a point-source and correcting for the interstellar
 extinction are $V=26.1\,\mbox{mag}$ and $B=26.2\,\mbox{mag}$,
 respectively. Values for E(B-V) were derived from dust maps
 (Schlegel et al.~1998). Converting these upper
 limits\footnote{Information on the conversion from the magnitude
 scale to the photon flux in Jy can be found at
 http://www.astro.utoronto.ca/$\sim$patton/astro/mags.htm} to a photon
 flux upper limit yields $1.322 \times 10^{-7} Jy$ for the V-band and
 $1.423 \times 10^{-7} Jy$ for the B-band, respectively.

\subsection{RADIO OBSERVATIONS OF \PSRB\label{radio_obs}}

 The ephemerides for the analysis of the X-ray data were obtained from
 radio observations and the measurement of pulse times--of--arrival
 (TOAs) using the 76-m Lovell radio telescope at Jodrell Bank
 Observatory. Table \ref{t:radio} summarizes the radio ephemerides of
 \PSRB. A dual-channel cryogenic receiver system sensitive to two
 orthogonal polarizations was used predominantly at frequencies close
 to 1400 MHz. The signals of each polarization were mixed to an
 intermediate frequency, fed through a multi-channel filter-bank and
 digitized.  The data were de-dispersed in hardware and folded
 on--line according to the pulsar's dispersion measure and topocentric
 period.  The folded pulse profiles were stored for subsequent
 analysis. In a later off--line processing step, every sub-integration
 spoiled by RFI was removed, the polarizations combined and the
 remaining sub-integrations averaged to produce a single
 total--intensity profile for the observation.  TOAs were subsequently
 determined by convolving, in the time domain, the averaged profile
 with a template corresponding to the observing frequency. The
 uncertainty on the TOA was found using the method described by Downs
 \& Reichley (1983), which incorporates the off--pulse rms noise and
 the `sharpness' of the template.  These TOAs were corrected to the
 solar system barycenter using the Jet Propulsion Laboratory DE200
 solar system ephemeris (Standish 1982). More details can be found in
 Hobbs et al.~(2004). Spectral data from \PSRB\, were obtained from
 the compilation of Maron et al.~(2000).

\begin{figure}
\centerline{\psfig{figure=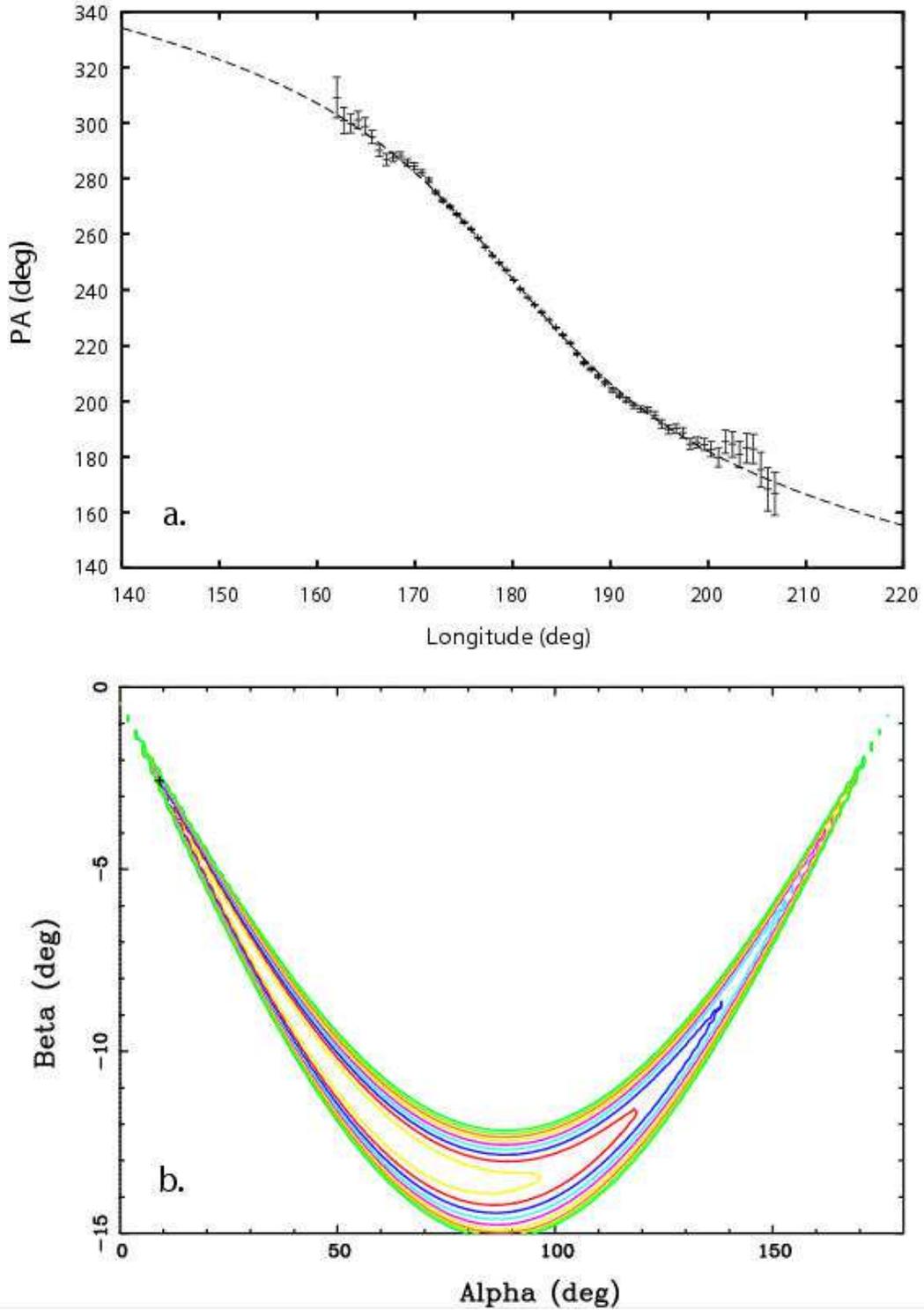,height=20cm,clip=}}
\caption[]{Fit of polarization data of PSR B0628-28. The upper panel {\bf a.}\,
shows the polarization position angle at 1400~MHz. The lower panel {\bf b.}\,
shows $\chi^2$-contours of the rotating vector model fit. The global maximum
is indicated by a + sign.  Alpha and Beta are the inclination and impact angles.} 
\label{polarisation}
\end{figure}

\begin{figure}
\centerline{\psfig{figure=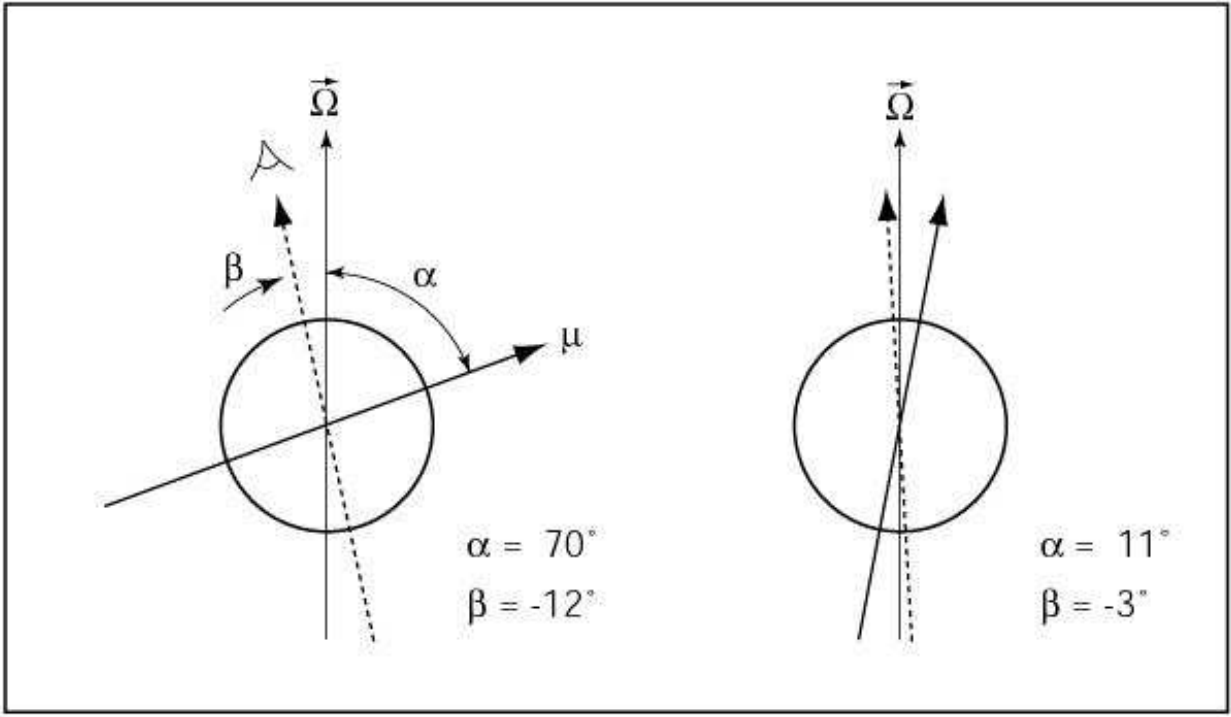,width=16.5cm,clip=}}
\caption[]{Possible emission beam geometries of \PSRB\, as deduced from fitting the
rotating vector model to the polarization angle swing observed at radio
frequencies. $\alpha$ is the inclination of the magnetic axis, $\beta$ the
minimum angle between the magnetic axis and the line of sight. $\vec{\Omega}$
is the rotation axis.} \label{PSRB_geometry}
\end{figure}

 The emission beam geometry of \PSRB\, is not known but can be
 estimated from fitting the canonical rotating vector model to the
 polarization position angle from radio data (e.g.~Lorimer \& Kramer
 2005). We used the polarization profiles observed at 408 MHz and 1400
 MHz by Gould \& Lyne (1998, see reference for details about observing
 set-up and calibration procedure) to determine the inclination of the
 magnetic axis to the rotation axis to be $\alpha=70^o$. The impact
 angle of the line of sight with the magnetic axis is fitted to be
 $\beta=-12^o$. As usual, the small duty cycle of 
 polarized emission available for modeling produces a distinct 
 correlation between the fit results for $\alpha$ and $\beta$. We 
 demonstrate this by showing the $\chi^2$-sphere for the 1400-MHz 
 least-squares fit in $\alpha$ and $\beta$ in Figure \ref{polarisation}. 
 Acceptable fits lie in the inside of the sickle shaped goodness of 
 fit contours. At the higher frequency of 1400~MHz shown here, the 
 formal global minimum of the least-squares fit resides at the edge 
 of the contours main body, in a small isolated point located at 
 $\alpha=11^o$ and $\beta=-3^o$ (cf.~Figure \ref{polarisation}).  In
 order to decide, which solution in this rather large solution space
 is physically more acceptable, we can make use of the pulse shape
 information: the measured 50\% width of the radio pulse amounts to
 $\sim 17^o$.  From studies of a large sample of radio pulsars, one
 finds a relationship between the beam radius and the spin period (see
 Lorimer \& Kramer 2005 and references therein). Based on such
 results, for a spin period of 1.244~s, one expects a beam width of
 $\sim 11^o$ for a line of sight cutting through the center of the
 emission cone (i.e.~$\beta \sim 0^o$).  An observed pulse width
 larger than this value may be a good indication of the possibility
 that the magnetic inclination angle is small, producing a path of the
 line-of-sight through the beam that is curved rather than
 straight. The fairly shallow gradient of the position angle curve may
 also add weight to this small-$\alpha$ solution, but a combination of
 the $\chi^2$-contours obtained for 408~MHz and 1400~MHz makes a
 nearly orthogonal solution $\alpha\sim 70^o, \beta\sim-12$ the most
 likely one. A sketch which illustrates possible beam geometries is 
 given in Figure \ref{PSRB_geometry}.

 \subsection{XMM-NEWTON OBSERVATIONS OF \PSRB \label{xray_obs}}

 \PSRB\, was observed with XMM-Newton\footnote{See http$://$xmm.vilspa.esa.es$/$
 for a description of XMM-Newton, its instrumentation and the various detector
 modes available for observations.} on April 20, 2004 (XMM rev.~773)
 for a total on-source time of $\sim 48\,500$s. We used the EPIC-PN
 camera as the prime instrument and operated both MOS1/2 cameras in
 PrimeFullWindow mode to obtain imaging and spectral data. The EPIC-PN
 camera was setup to operate in PrimeLargeWindow readout mode which
 provides imaging, spectral and timing information with a temporal
 resolution of 48 ms. The higher temporal resolution of the EPIC-PN
 small-window mode would have been attractive but an overkill in view
 of its $\sim 30\%$ higher dead-time (see e.g.~Becker \& Aschenbach
 2002 for a summary of XMM-Newton instrument modes suitable for pulsar
 studies). The medium filter was used for the MOS1/2 cameras and the
 thin filter for the EPIC-PN. The expected lower efficiency of the RGS
 means that it is of limited use for the given exposure time. The
 optical field is well known from our NTT/EMMI observations so that we
 do not report on the analysis of XMM's optical monitor (OM)
 data. This data, which were taken with the OM in standard
 configuration are superseded in sensitivity by our ground based
 ESO/NTT observation (see \S \ref{optical_obs}).  A summary of
 exposure times, instrument modes and filters used for the X-ray
 observation is given in Table ~\ref{t:xray_obs}.

 XMM-Newton data have been seen to show timing discontinuities in the
 photon arrival times with positive and negative jumps of the order of
 one to several seconds (Becker \& Aschenbach 2002; Kirsch et
 al.~2003). Inspecting the log-files from our processing of raw data
 we found that the EPIC-PN data of \PSRB\, exhibit a clock
 discontinuity showing one negative jump of 15s. We therefore used 
 a release track version of the XMM-Newton SAS (Standard Analysis Software) 
 for the analysis of the EPIC-PN data. This software detects and corrects 
 most of the timing discontinuities during data processing. In addition, known 
 timing offsets due to ground station and space craft clock propagation 
 delays are corrected by this software in using new reconstructed time 
 correlation (TCX) data. Barycenter correction of the EPIC-PN data and 
 all other analysis steps were performed by using SAS Version 6.1.

 Data screening for times of high background was done by inspecting
 the light-curves of the MOS1/2 and PN data at energies above 10
 keV. Apart from having a rather high sky background contribution,
 strong X-ray emission from soft proton flares are seen at various
 times during the observation. Creating the light-curves with bins of
 100s, we rejected those bins where the MOS1/2 light-curves had more
 than 130 cts/bin. For the EPIC-PN data we rejected times with more
 than 90 cts/bin (for comparison, data with a typical low sky
 background require the rejection of events at a level of $\sim10$
 cts/bin).  The data screening reduced the effective exposure time for
 the MOS1/2 to 35.6 ksec and 34.6 ksec, respectively. For the EPIC-PN
 data the effective exposure time was reduced to 31.7 ksec. The net
 exposure time after rejecting times of high sky background thus is
 only about 74\% of the requested observing time.

 For the spectral analysis based on the MOS1/2 data we used only those
 events with a detection {\em pattern} between $0-12$ (i.e. single,
 double and triple events) and the {\em flag} parameter set to less
 than or equal to 1. The latter criterion excludes events which are
 located near to a hot pixel, or to a bright CCD column, or which are
 near to the edge of the CCD. For the EPIC-PN timing and spectral
 analyzes, we used only single and double events, i.e.~those which
 have a pattern parameter of less than, or equal to, 4 and a flag
 parameter equal to zero. The energy range of the MOS1/2 and EPIC-PN
 CCDs was restricted to $0.2-10$ keV for the spectral and timing
 analysis.

 \section{ANALYSIS OF THE XMM-DATA OF \PSRB\label{PSRB}}

 The putative X-ray counterpart of \PSRB\, is detected with high
 significance in both the MOS1/2 and EPIC-PN data. The XMM-Newton net
 counting rates within the $0.2-10$ keV band are $0.0047\pm 0.0008$
 cts/s (MOS1), $0.0042 \pm 0.0008$ cts/s (MOS2), and $0.021 \pm 0.003$
 cts/s (EPIC-PN), respectively. A maximum-likelihood source-detection
 did not yield any evidence for a spatial extent of \RX\, of larger
 than 15 arcsec, corresponding to the HEW (Half Energy Width) of the 
 instruments' point spread function.

 \subsection{Timing Analysis\label{PSRB_timing}}

 We used the EPIC-PN large-window mode data for the timing
 analysis. Given the pulsar period of $\sim 1.244$s the temporal
 resolution of 48ms allows us to perform a detailed timing analysis in
 searching for X-ray pulsations and to construct a pulse profile with
 up to $\sim 26$ independent phase bins, if supported by the photon
 number statistics.

 Events were selected from a circle of 25 arcsec radius centered on
 \RX. This extraction region contains 80\% of the point source
 flux. For the barycenter correction we applied the standard
 procedures for XMM-Newton data using {\em barycen-1.17.3}\/ and the
 JPL DE200 Earth ephemeris (Standish 1982) to convert photon arrival
 times from the spacecraft to the solar system barycenter (SSB) and
 the barycentric dynamical time (TDB). The pulsar radio timing
 position (cf.~Table \ref{t:radio}) was used for the barycenter
 correction. The spin-parameters $f$ and $\dot{f}$ of \PSRB\, are
 known with high precision from our contemporaneous radio
 observations, covering the mean epoch MJD=53063.4104692111723
 (TDB@SSB) of the XMM-Newton observation. \PSRB\, is not known to show
 timing irregularities (glitches) so that we can fold the photon
 arrival times using the pulsar's radio ephemeris (see Table
 \ref{t:radio}).
 The statistical significance for the presence of a periodic signal
 was obtained from a $Z^2_n$-test with $1-10$ harmonics in combination
 with the H-Test to determine the optimal number of harmonics (De
 Jager 1987; Buccheri \& De Jager 1989). The optimal number of phase
 bins for the representation of the pulse profile was determined by
 taking into account the signal's Fourier power and the optimal number
 of harmonics deduced from the H-Test (see Becker \& Tr\"umper 1999
 and references therein).

 Within the $0.2-10$ keV energy band, 1290 events were available for
 the timing analysis of which $\sim 35\%$ are estimated to be
 background. The $Z^2_n$-test gave 66.1 for $n=4$ harmonics
 ($Z^2_1=48.41$). According to the H-Test, the probability of
 measuring $Z^2_4=66.1$ by chance is $\sim 3 \times 10^{-11}$ thus
 confirming \RX\, (\CXC, resp.) to be the counterpart of \PSRB\, and
 establishing it firmly to be an X-ray pulsar!

\begin{figure}
\centerline{\psfig{figure=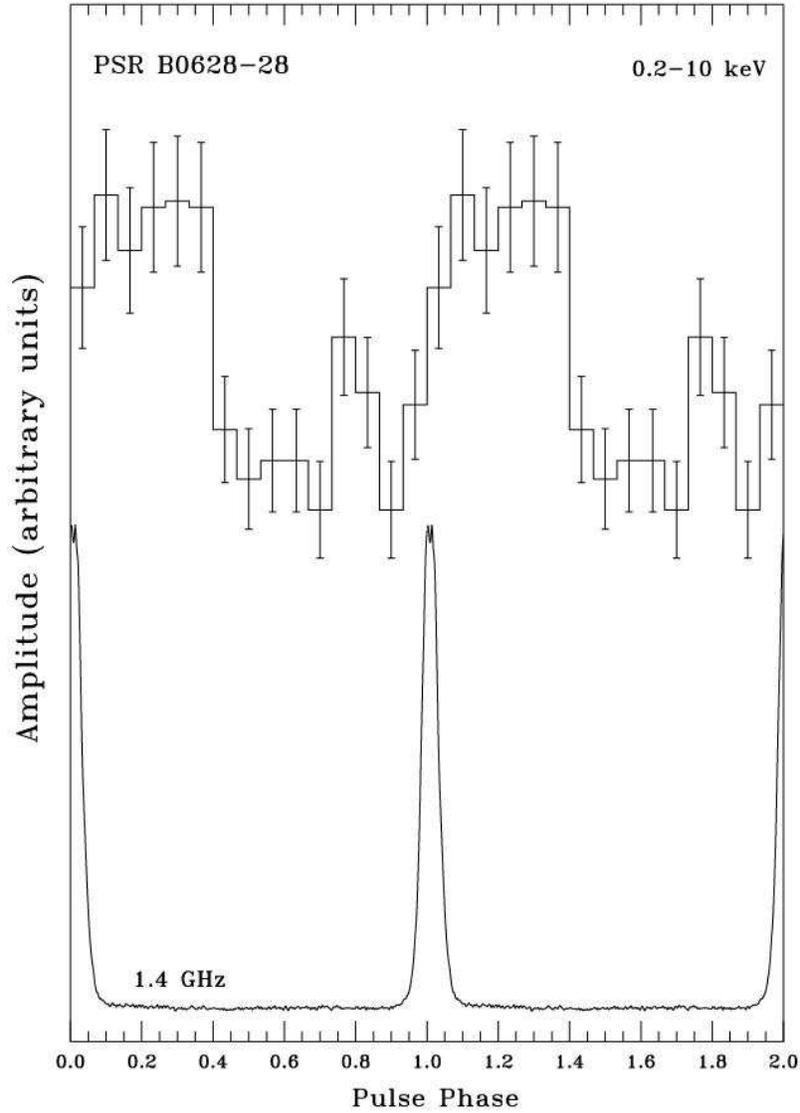,height=15cm,clip=}}
\caption[]{Integrated pulse profiles of \PSRB\, as observed in the $0.2-10$ keV
band (top) and and at 1.4 GHz with the Jodrell Bank radio observatory (bottom).
X-ray and radio profiles are phase related. Phase zero corresponds to the mean
epoch of the XMM-Newton observation. Two phase cycles are shown for clarity.}
\label{PSRB_x_radio_profiles}
\label{PSRB_pulseprofiles}
\end{figure}

 Figure \ref{PSRB_pulseprofiles} depicts the X-ray pulse profile of
 \PSRB\, for the $0.2-10$ keV energy band.  As indicated by the higher
 harmonic content and as can be seen easily in the figure, the pulse
 shape is not sinusoidal but double peaked, with a main broad peak of 
 width $\sim 180^\circ$ and a narrow pulse component of $\sim 45^\circ$
 width longitude. By taking the center of mass of both pulse components as 
 a reference point, the separation in phase between both peaks is $\sim 0.4$ 
 ($144^\circ$ longitude). The fraction of pulsed events in the 
 $0.2-10$ keV energy range is determined to be of $39 \pm 6 \%$ by 
 using a bootstrap method (Swanepoel, de Beer \& Loots 1996; Becker 
 \& Tr\"umper 1999). Restricting the timing analysis to the $0.2-1.0$ 
 keV and  $1.0-2.1$ keV energy bands yields the pulsed fractions 
 $41 \pm 7\%$ and $42 \pm 10\%$, respectively, which do not indicate 
 a significant energy dependence over the $0.2-10$ keV bandpass, 
 although the decreasing signal to noise ratio beyond $\sim 2.1$ keV  
 makes any conclusion for this energy band merely tentative. 

 Comparing the X-ray pulse profile with the one taken with the Jodrell
 Bank radio telescope at 1.4 GHz shows that both profiles are markedly 
 different. Measuring the TOAs of the radio and X-ray pulses shows that 
 the narrow X-ray pulse leads the radio pulse by $\sim 0.2$ ($72^\circ$  
 longitude) in phase. The radio pulse itself leads the broader X-ray 
 pulse by about the same amount of $72^\circ$ longitude. The arrival of 
 the radio pulse thus appears to be in the middle between the arrival 
 of the two X-ray pulse components.

 We note that uncertainties of the XMM-Newton clock against UTCs are not 
 relevant as those are on a scale of $\sim 100\,\mu s$ (Becker et al.~2005, 
 in prep.) and thus are a factor of $\sim 1000$ smaller than the bin width  
 of the X-ray pulse profile shown in Figure \ref{PSRB_x_radio_profiles}.

 \subsection{Spectral Analysis\label{PSRB_spectral}}

 The energy spectrum of \PSRB\, was extracted from the MOS1/2 data by
 selecting all events detected in a circle of radius 50 arcsec
 centered on the pulsar position. Using the XMM-Newton/EPIC-MOS model
 point spread function, 90\% of all events of a point source are
 within this region. The background spectrum was extracted from an
 annulus of outer radius 85 arcsec surrounding \PSRB. For the EPIC-PN
 data we used an extraction radius of 31 arcsec centered on \PSRB.
 This selection region includes $\sim 85\%$ of the point source
 flux. As the instrument focus is relatively close to the edge of the
 PN-CCD, we extracted the background spectrum from a source free
 region about one arc-minute east of the pulsar. Out-of-time events
 prevent us from extracting the background spectrum from a region
 located below the source and along the CCD read-out direction.

 In total, the extracted spectra include 814 source counts from the
 EPIC-PN camera and 430 source counts from the EPIC-MOS1/2
 detector. The spectral data were dynamically binned so as to have at
 least 30 counts per bin. Model spectra were then simultaneously fit
 to both the PN and MOS1/2 data.

 Amongst the single component spectral models, a power law model was
 found to give the statistically best representation (\chisq =34.6 for
 37 dof) of the observed energy spectrum. A single blackbody model
 did not give acceptable fits (\chisq =65.1 for 37 dof) and can be 
 rejected.  The power law model yields a column absorption of
 $N_H=6.0_{-1.8}^{+3.1}\times 10^{20}\,\mbox{cm}^{-2}$, a photon-index
 $\alpha = 2.63_{-0.15}^{+0.23}$ and a normalization of
 $1.0_{-0.1}^{+0.13}\times 10^{-5}$ photons cm$^{-2}$ s$^{-1}$
 keV$^{-1}$ at $E=1$ keV. The errors represent the $1\sigma$
 confidence range for one single parameter of interest. 
 For the unabsorbed energy flux we measured $f_x=
 3.36_{-0.17}^{+0.22}\times 10^{-14}\, \,{\rm ergs\,\, s}^{-1}\,{\rm
 cm}^{-2}$ in the $0.5-10$ keV band, yielding an X-ray luminosity of
 $L_x=8.42_{-0.54}^{+0.42} \times 10^{30}\,{\rm ergs\,\, s}^{-1}$ for
 a distance of 1.45 kpc.  For the ROSAT energy band, $0.1-2.4$ keV, we 
 measured the flux to be $f_x=9.4_{-2.3}^{+5.0}\times 10^{-14}\,\,
 {\rm ergs\,\, s}^{-1}\,{\rm cm}^{-2}$, yielding an X-ray luminosity 
 of $L_x=2.4_{-0.6}^{+1.25}\times 10^{31} \, {\rm ergs\,\, s}^{-1}$.  
 These luminosities imply the huge rotational energy to X-ray energy
 conversion factors $L_x/\dot{E}= 58.3\times 10^{-3}$ within $0.5-10$
 keV and $163.4\times 10^{-3}$ if transformed to the ROSAT band (see
 \S\ref{discussion} for a discussion).

\begin{figure}
\centerline{\psfig{figure=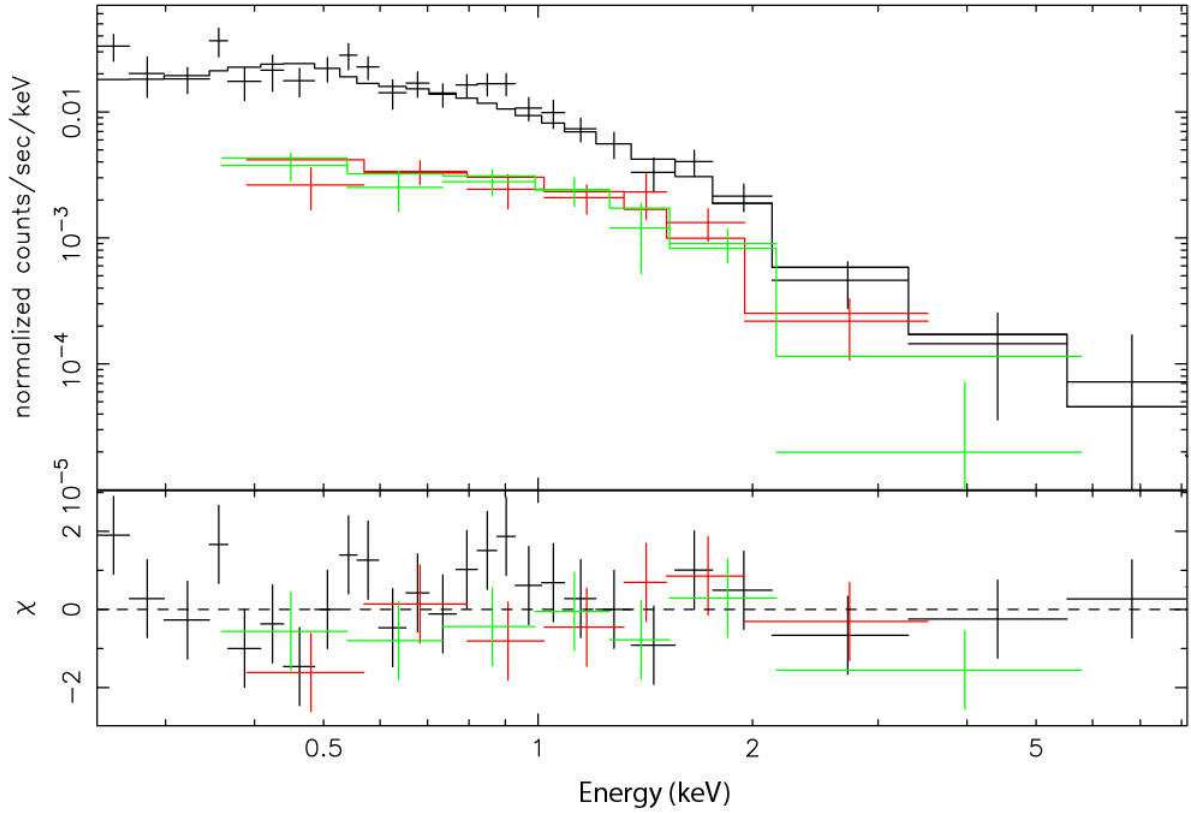,width=16cm,clip=}}
\caption[]{Energy spectrum of \PSRB\, as observed with the EPIC-PN (upper spectrum)
and MOS1/2 detectors (lower spectra) and simultaneously fitted to an absorbed power
law model ({\it upper panel}) and contribution to the \chisq\, fit statistic
({\it lower panel}).} \label{PSRB_pl_spectrum}
\end{figure}

\begin{figure}
\centerline{\psfig{figure=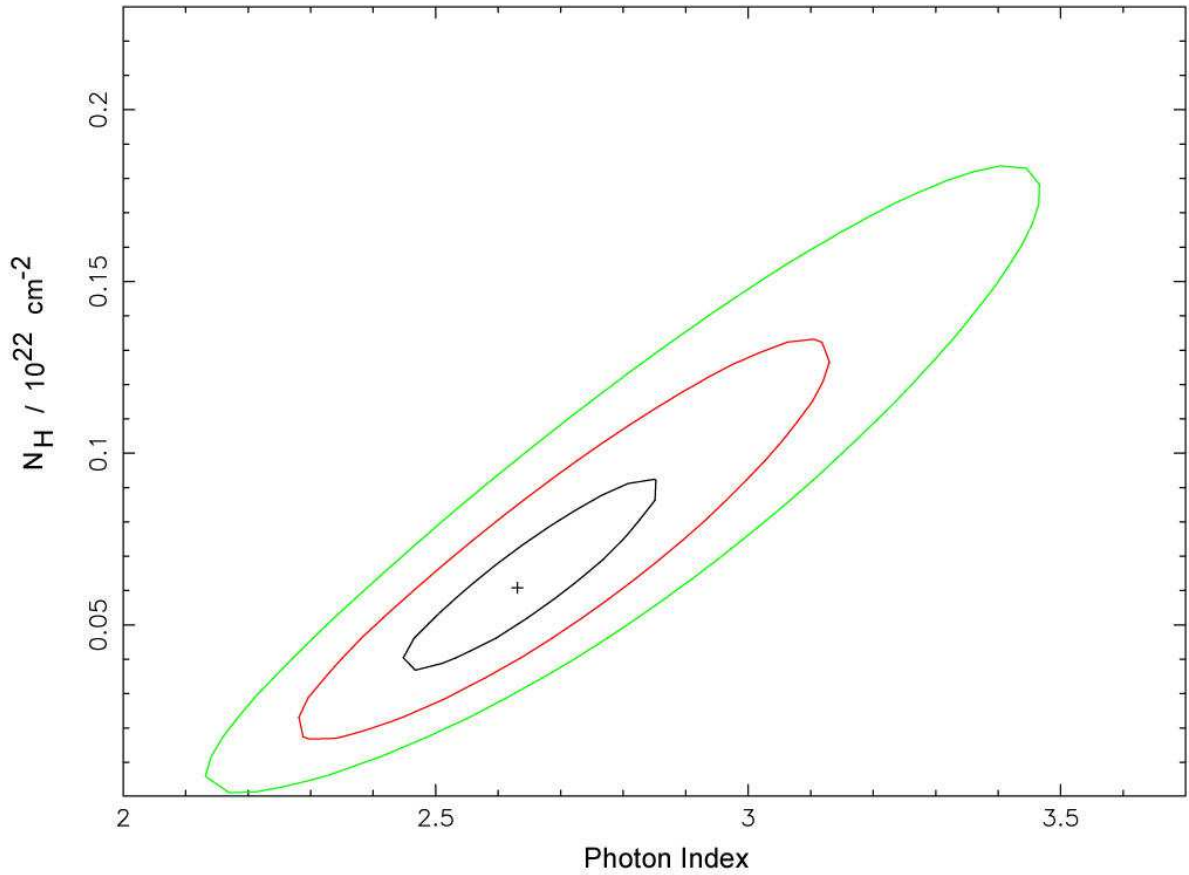,width=16cm,clip=}}
\caption[]{Contour plot showing the relative
parameter dependence of the photon index vs.~column absorption for the power law
fit to the \PSRB\, data. The three contours represent the $1-\sigma$, $2-\sigma$
and $3-\sigma$ confidence contours for one parameters of interest. The `+' sign
marks the best fit position, corresponding to $\chi^2_{min} =0.8650$ for 40 dof.}
\label{PSRB_pl_contour}
\end{figure}

 The best-fit power law spectrum and residuals are shown in
 Figure~\ref{PSRB_pl_spectrum}.  Contour plots showing the relationship 
 between the photon index and the column absorption for various confidence 
 levels are shown in Figure \ref{PSRB_pl_contour}. 

 Testing composite spectral models consisting of two Planckian components 
 or of a Planckian plus power law component resulted in fits which had a 
 chi-square of 28.6 (for 35 dof), i.e.~a goodness comparable to what we found 
 in the single component power law fit. The F-test statistic for adding the 
 extra blackbody spectral component to the power law model yields a 
 probability of only 97.2\%, i.e.~slightly more than $2\sigma$. The 
 justification to include a thermal component to the power law spectrum 
 thus is not particularly strong by statistical means. 

 The parameters fitted for the composite Planckian plus power law model are a 
 column absorption of $N_H \le 2.5 \times 10^{20}\,\mbox{cm}^{-2}$,
 a photon-index $\alpha = 2.27^{+0.23}_{-0.13}$ and a normalization of the power
 law component of $4.8^{+1.6}_{-0.9} \times10^{-6}$ photons cm$^{-2}$ s$^{-1}$
 keV$^{-1}$ at $E=1$ keV.  The blackbody temperature and the size of the projected
 emitting area are $kT = 0.25_{-0.04}^{+0.05}$ keV and $R_{bb}= 69_{-25}^{+30}$\,m, 
 assuming a pulsar distance of 1.45 kpc. For the unabsorbed energy flux and 
 luminosity we compute $f_x = 2.8^{+3}_{-1} \times 10^{-14} \,\mbox{erg s}^{-1}\mbox{cm}^{-2}$
 and $L_x = 7.0^{+8}_{-2}  \times 10^{30}\,\mbox{erg s}^{-1}$ for the $0.5-10$ keV band. For 
 the $0.1-2.4$ keV energy band we compute the model flux to be 
 $f_x = 4.0^{+6.0}_{-1.7} \times 10^{-14} \,\mbox{erg s}^{-1}\mbox{cm}^{-2}$, yielding an 
 X-ray luminosity of   $L_x = 1.0^{+1.5}_{-0.5}  \times 10^{31}\,\mbox{erg s}^{-1}$.
 The rotational energy to X-ray energy conversion factors obtained from this combined 
 model are $L_x/\dot{E}= 48.4\times 10^{-3}$ within $0.5-10$ keV and $69.8 \times 10^{-3}$ 
 if transformed to the ROSAT band.

\begin{figure}
\centerline{\psfig{figure=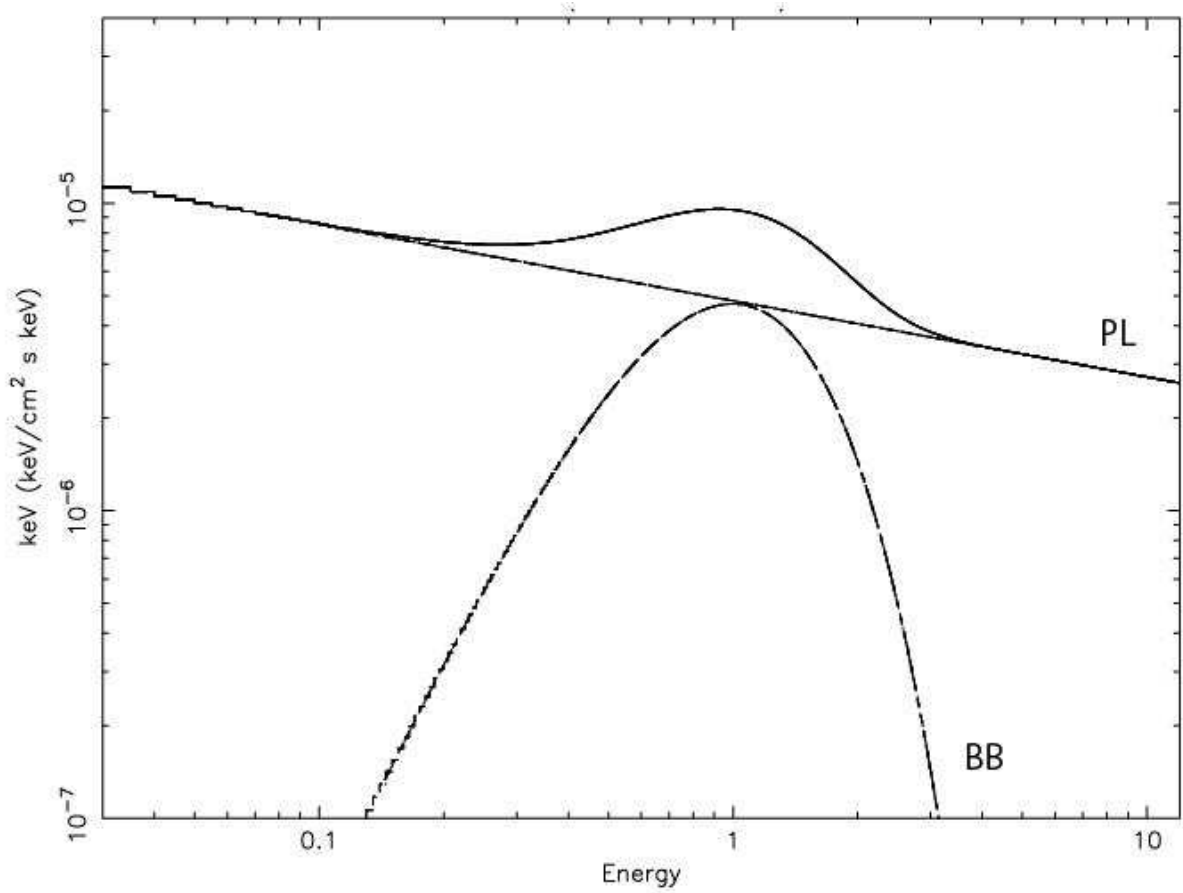,width=16cm,clip=}}
\caption[]{Blackbody plus power law spectral components and combined model as fitted to
the spectral data of \PSRB.}
\label{PSRB_bb_pl_model}
\end{figure}

 In computing the relative contributions of the thermal and non-thermal spectral components
 we find that for the best fitting parameters $\sim 20\%$ of the X-ray flux within the 
 $0.1-2.4$ keV band could be of thermal origin and emitted from heated polar caps. In
 stretching the errors to the limits, however, a maximum of up to $\sim 100\%$ thermal 
 flux can not be excluded for this energy band. At higher and lower energies non-thermal 
 emission will dominate though. Figure \ref{PSRB_bb_pl_model} illustrates the relative 
 contributions of the thermal and non-thermal spectral components as indicated by the best 
 fitting model parameters. 

 Defining the size of a presumed polar cap as the foot points of the neutron star's dipolar 
 magnetic field, the radius of the polar cap area is given by $\rho=\sqrt{2\pi R^3/c P}$ with 
 $R$ being the neutron star radius, $c$ the velocity of light and P the pulsar rotation period 
 (see e.g.~Michel 1991).  For \PSRB\, with a rotation period of 1.244 s this yields a polar cap 
 radius of $\rho\sim 130$ m which is not too different from the size of the blackbody emitting 
 area fitted in the composite blackbody plus power law model for an assumed distance of 1.45 kpc. 

 The two component blackbody model would yield an alternative description of the spectrum as well, 
 based on thermal emission mechanisms only. The parameters fitted by this model are a column 
 absorption of $N_H \le 0.9 \times 10^{20}\,\mbox{cm}^{-2}$, blackbody temperatures of 
 $\mbox{kT}_1=0.142_{-0.01}^{+0.22}$ keV, $\mbox{kT}_2=0.402_{-0.04}^{+0.09}$ keV, and radii of
 the projected emitting areas of $R_1=278.8_{-59.2}^{ +116.4}$\,m, $R_2=34.8_{-11.33}^{+9.75}$\,m.

\begin{figure}
\centerline{\psfig{figure=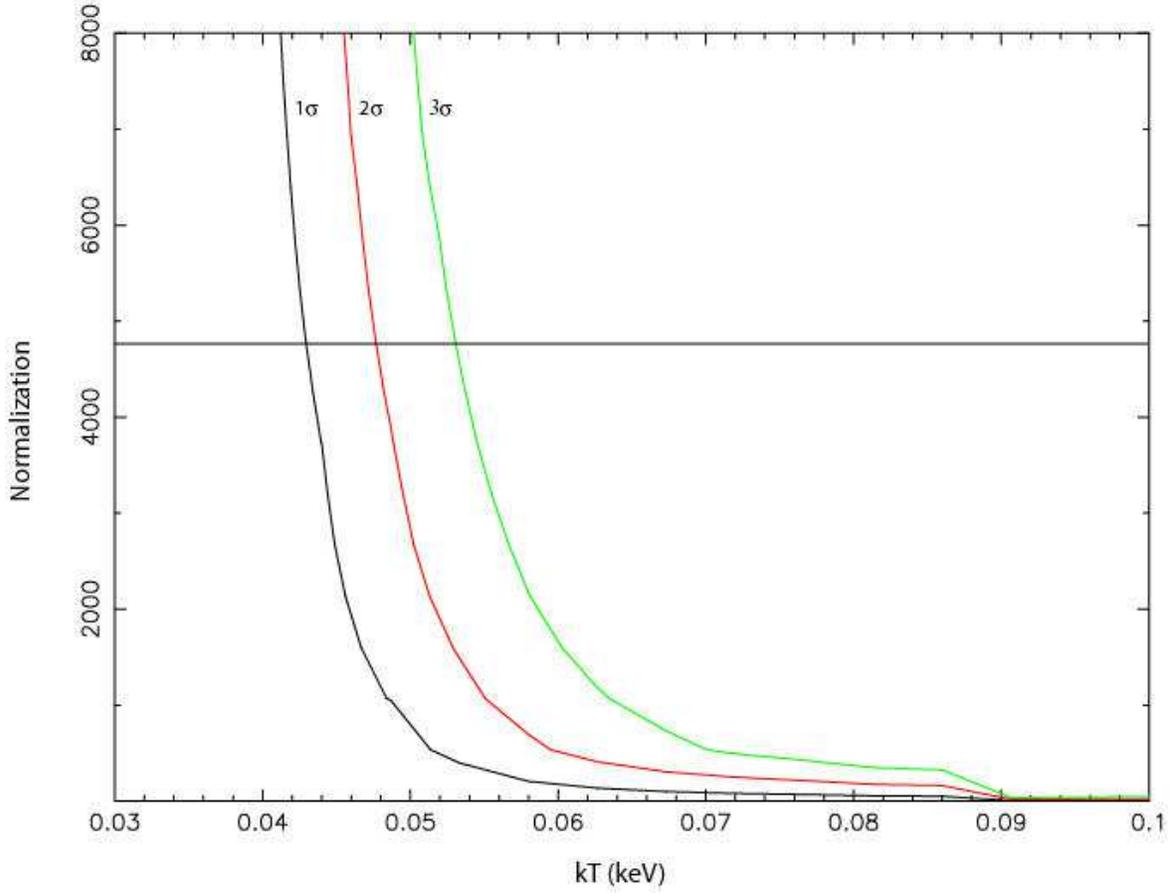,width=16cm,clip=}}
\caption[]{Portion of the confidence contours showing the blackbody normalization
versus blackbody temperature for the composite Planckian plus power law model  
(see text). The horizontal line at a normalization of 4772 corresponds to a neutron 
star radius of 10 km at a pulsar distance of 1.45 kpc. The contours correspond to 
$\chi^2_{min}=33.91$ plus 2.3, 6.17 and 11.8 which are the $1-\sigma$, $2-\sigma$ 
and $3-\sigma$ confidence contours for 2 parameters of interest.} 
\label{PSRB_cooling_pl_contour}
\end{figure}

 With a spin-down age of $\sim 2.7\times 10^6$ years \PSRB\, should
 still have some residual heat content from its birth event. Depending
 on the equation of state the surface temperature could be in the
 range $\sim 1-3 \times 10^5$ K (cf.~Becker \& Pavlov 2002 and
 references therein) and then would contribute on a low level to the
 detected soft X-ray emission. Clearly, the power law spectral model
 fit does not require this extra thermal component by statistical
 means but to estimate the upper limit for the surface temperature of
 \PSRB\, we have added a blackbody component to the best fitting power
 law model.  We then calculated the confidence ranges of the blackbody
 normalization and temperature by leaving all other parameters
 free. The resulting contours, computed for two parameters of
 interest, are shown in Figure \ref{PSRB_cooling_pl_contour}. For the
 thermal emission to be emitted from the whole surface of a neutron
 star of radius 10 km we find a $2\sigma$ surface temperature upper
 limit of $T_s^\infty < 5.3 \times 10^5$ K which is somewhat above the
 temperatures predicted by cooling models (e.g.~Page \& Applegate
 1992; Yakovlev et al.~1999).

 The spectral parameters of the various models fitted to the energy spectrum 
 of \PSRB\, are summarized in Table \ref{spectral_fits_J0628}.

\subsection{Multi-wavelength Spectrum\label{PSRB_multi_spec}}

\begin{figure}
\centerline{\psfig{figure=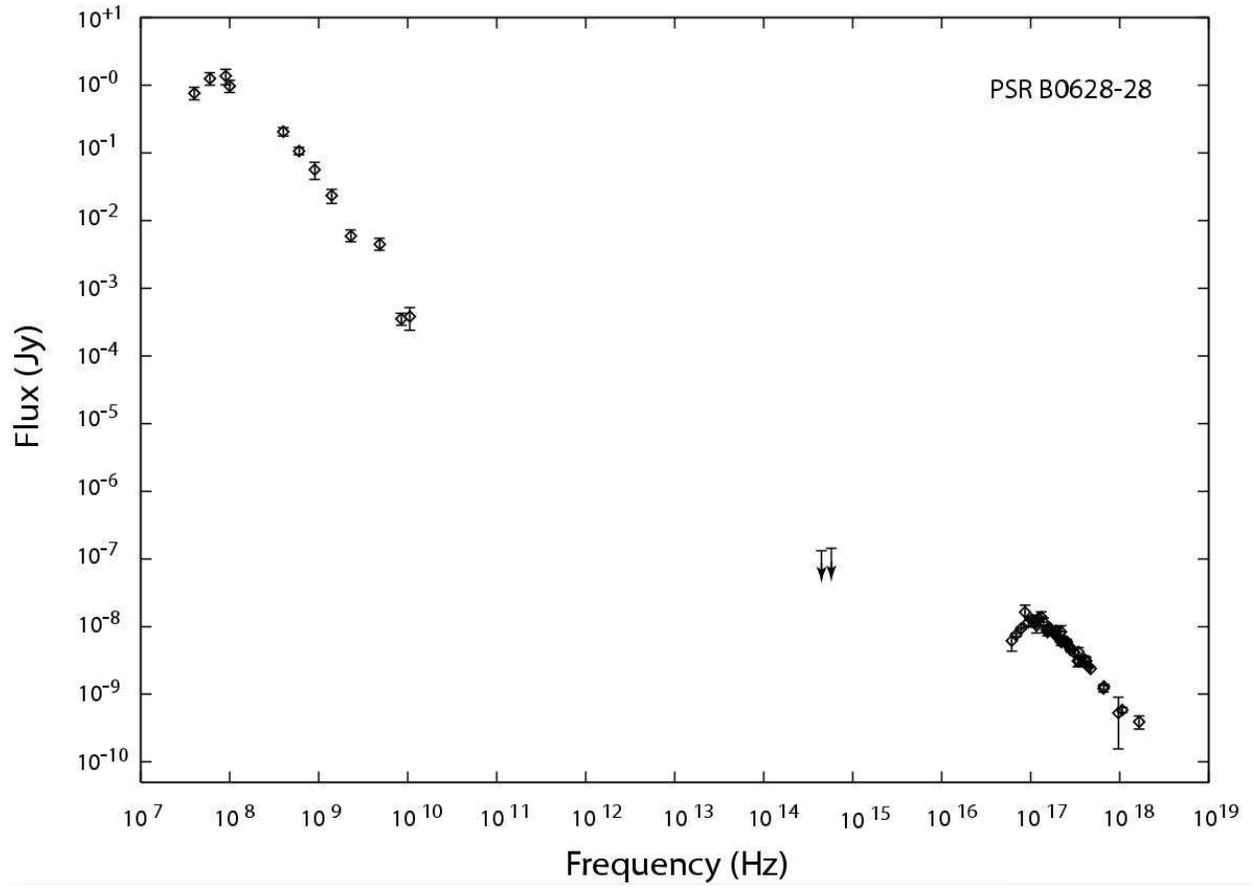,width=16.5cm,clip=}}
\caption[]{Combined radio, optical and X-ray spectral data of \PSRB.}
\label{PSRB_broadband_spectrum}
\end{figure}

 In order to construct a broadband spectrum combining all spectral
 information available from \PSRB\, we adopted the radio spectrum from
 Maron et al.~(2000) and plotted it in Figure
 \ref{PSRB_broadband_spectrum} together with the optical V- and B-band
 upper limits from our ESO/NTT observations and the XMM-Newton
 observed pulsar spectrum.  The radio spectrum has been measured up to
 10.6~GHz and can be modeled by a simple power-law with an average
 photon index of $-2.9 \pm 0.1$ (Maron et al. 2000), even though
 in contrast to measurements reported by Reyes et al.~(1995), the
 spectrum does show a turn-over at low frequencies (see Maron et
 al.~2000 and references therein). The flux density in the radio part of
 the spectrum, which is supposed to be due to coherent radiation, is
 several orders of magnitude greater than the extrapolated optical or
 X-ray flux densities.

 Extrapolating the power law spectrum which describes the XMM-Newton
 data to the optical V- and B-bands yields a photon flux which exceeds
 the measured upper limits by more than an order of magnitude. This
 suggest that the broadband spectrum, if entirely non-thermal, has to
 break somewhere before or in the soft channels of the X-ray
 spectrum. To test this hypothesis we have fitted a broken power law
 model to the XMM-Newton data and found it providing a better
 description (\chisq = 27.576 for 37 dof) of the observed X-ray
 spectrum than the single power law model does. A broken power law
 model which is in agreement with the optical V- and B-band upper
 limits and the XMM-Newton data has its break point fitted at
 $E_{break}= 0.85^{+0.13}_{-0.16}\,\mbox{keV}$. The photon-index for
 $E < E_{break}$ and $E > E_{break}$ is found to be $\alpha_1=
 1.45^{+0.04}_{-0.003}$ and $\alpha_2 =2.71^{+0.34}_{-0.26}$,
 respectively, with a normalization of $1.6 \times10^{-5}$ photons
 cm$^{-2}$ s$^{-1}$ keV$^{-1}$ at 1 keV. The column absorption is
 compatible with an upper limit of $N_H=1.5\times
 10^{20}\,\mbox{cm}^{-2}$.  The errors represent the $1-\sigma$
 confidence range calculated for two parameters of interest.

 \section{DISCUSSION \& SUMMARY\label{discussion}}

 We have investigated the optical and X-ray emission properties of 
 \RX\, and identified it to be the X-ray counterpart of the old 
 rotation-driven pulsar \PSRB\, by detecting its X-ray pulses at the 
 radio pulsar's spin frequency. Its X-ray pulse profile is characterized by 
 a single broad peak and a second narrow pulse component which leads the 
 main pulse by $\sim 144^\circ$.  A comparison of the pulse profiles
 observed in the $0.2-10$ keV band and at 1.4 GHz shows that the  
 X-ray pulse profile is markedly different from the single narrow
 peaked profile observed in the radio band. The XMM-Newton observed 
 pulsed fraction is  $(39\pm 6)\%$ with no strong energy dependence  
 in the $0.2-10$ keV bandpass. The pulsar's X-ray spectrum is very well 
 described by a power law spectrum, indicating that non-thermal radiation  
 processes dominate the pulsar's X-ray emission in the $0.2-10$ keV band. 
 A $\sim 20\%$ thermal flux contribution from heated polar caps is inferred 
 from the best fitting parameters of a combined power law plus Planckian 
 spectrum.  

 The emission beam geometry of \PSRB\, is estimated from radio polarization 
 data taken at 408 MHz and 1400 MHz. A formal best fit of the 1400 MHz data 
 yields $\alpha\sim 11^\circ$ for the inclination of the magnetic axis
 to the rotation axis and $\beta\sim -3^\circ$ for the impact angle of the
 line of sight. A combination of results obtained from 408 MHz and 1400 MHz
 data, however, makes a nearly orthogonal solution with $\alpha\sim 70^\circ$ 
 and $\beta\sim -12^\circ$ the most likely one. We would like to point out,
 however, that in the case the beam geometry would be $\alpha\sim 11^\circ$,
 $\beta\sim -3^\circ$ (which currently can not be ruled out) it would disclose 
 us a full view of the pulsar polar cap.

 Comparing the pulsar's X-ray luminosity, $L_x(\mbox{0.1-2.4 keV}) =
 2.36_{-0.57}^{+1.25}\times 10^{31} \,{\rm ergs\,\, s}^{-1}$, as inferred
 by the power law spectral model and estimated under the assumption of a 
 pulsar distance of 1.45 kpc, with the pulsar's spin-down energy, $\dot{E}= 
 1.445\times 10^{32}\,\mbox{erg/s}$, we find that \PSRB\, emits the huge 
 amount of $\sim 16\%$ of its spin-down energy into the soft X-ray band. If
 confirmed, \PSRB\, would be the first X-ray over-luminous rotation
 powered pulsar identified among all $\sim 1400$ radio pulsars known
 today. It is not known what could cause such a high X-ray luminosity. Among 
 the most possible scenarios are extreme re-heating from vortex creep, 
 accretion from the ISM, decay of the pulsar magnetic field, or extreme
 polar cap heating (cf.~Harding \& Muslimov 2001,2002), though all this 
 is not observed so far in any of the other X-ray detected rotation 
 powered pulsars. 

 Well known pulsars which, according to their timing parameters, fall
 into the same category as \PSRB\, are PSRs B1929+10, B0950+08,
 B0823+26 and J2043+2740. All four have been detected by ROSAT and
 ASCA at the limits of their sensitivity and have been investigated
 using XMM-Newton recently (Becker et al.~2004; 2005), but none of 
 them shows evidence for the X-ray over-luminosity observed in \PSRB. 
 Thus, is \PSRB\, unique among all known rotation powered pulsars? At 
 least its observed spectral and temporal emission
 properties do not support this view.  The results found for PSRs
 B1929+10, B0950+08, B0823+26 and J2043+2740 are all in line with the
 emission properties observed in \PSRB. As for \PSRB, the X-ray
 emission from these old pulsars seems to be  dominated by 
 non-thermal radiation processes. None of the observed spectra 
 {\em required} adding a thermal component consisting of either a hot 
 polar cap or surface cooling emission to model the data. The X-ray 
 spectrum of PSR B0950+08 is best described by a single power law of 
 photon-index $\alpha=1.93^{+0.14}_{-0.12}$. Its X-ray emission is 
 pulsed with two peaks per period and a phase separation of $\sim 144^\circ$ 
 between the two pulse components. Its pulsed fraction is $(28 \pm 6)\%$ in
 the $0.2-10$ keV band. The spectral and temporal emission properties
 observed from PSR B1929+10 and PSR B0823+26 are very
 similar to those seen in PSR B0950+08 and \PSRB (Becker et al.~2004). 
 Their energy spectra can be adequately described by a single power law with 
 photon-index
 $\alpha = 2.72^{+0.11}_{-0.09}$ and $\alpha=2.5^{+0.9}_{-0.45}$,
 respectively. Fitting combined Planckian plus power law models
 to the observed spectral data from these pulsars, however, is possible
 as well and indicates thermal contributions from heated polar caps
 to the soft X-ray bands of the order of $\sim 10-40\%$ according to
 the best fitting model spectra. The fraction of pulsed photons for 
 all these pulsars are in the range of $\sim 30-50\%$. 
 
 The similarity in the emission properties to other X-ray detected
 pulsars from the same class does not support the assumption that
 there is anything special in the emission properties of \PSRB\, which
 justifies the high spin-down to X-ray energy conversion. The
 spin-down to X-ray energy conversion efficiency for the sample of
 ROSAT detected rotation-powered pulsars was investigated by Becker \&
 Tr\"umper (1997) who described the available data with the relation
 $L_x\sim 10^{-3}\,\dot{E}$. This relation is corrected for possible
 thermal contributions e.g.~in cooling neutron stars and was found to
 fit the data within the $0.1-2.4$ keV band assuming isotropy. Arguing
 that the $2-10$ keV energy band should be better suited for this
 study because any thermal contribution and spectral fitting
 uncertainties from interstellar absorption is minimized, Possenti et
 al.~(2002) performed a similar analysis using ROSAT, ASCA and
 BeppoSAX data of a somewhat larger sample than was available
 before. For faint sources like the old and nearby radio pulsars and
 most millisecond pulsars, Possenti et al.~(2002) obtained the flux in
 the $2-10$ keV band by simply upscaling the ROSAT results from Becker \& 
 Tr\"umper(1997). Those
 results, however, are not very meaningful, especially if there is no
 spectral information available and a simple counts-to-energy
 conversion requires the assumption of a spectral shape, as it was the
 case for the sample of old non-recycled pulsars.  Here, PSR B1929+10
 was taken as a prototype and X-ray fluxes were obtained by assuming
 that the X-ray emission from old pulsars originates entirely from 
 heated polar caps.  Our new spectral results obtained for \PSRB\, 
 and for the other old rotation-powered pulsars in recently XMM-Newton
 observations (cf.~Becker et al.~2004; 2005), however, shows that this
 assumption is no longer valid.

 Consequently, we do not follow their conclusions e.g.~of maximum 
 X-ray emissivity but assume that the conversion efficiency $L_x/ \dot{E}\,$ 
 for \PSRB\, within $0.1-2.4$ keV is in the same range of 
 $x=(0.5 - 5)\times 10^{-3}$ than observed by Becker et al.~(2004; 2005) 
 in PSR B1929+10 ($3.44\times 10^{-3}$), B0950+08 ($2.85\times 10^{-4}$), 
 B0823+26 ($5.1\times 10^{-4}$), and J2043+2740 ($5\times 10^{-4}$) using 
 XMM-Newton. We are therefore in favor of the interpretation that in
 $L_x=4 \pi d^2_{DM} f_x$ the 
 distance of 1.45 kpc, based on the radio dispersion measure and the Galactic 
 free electron density model NE2001 (Cordes \& Lazio 2002), is overestimated. 

 If we take the errors in the X-ray flux measured from \PSRB\, into
 account, we obtain a pulsar distance $d(x)=\sqrt{\dot{E} x / 4 \pi
 f_x}$ which is about 5 to 22 times closer than the radio dispersion
 measure inferred distance.  Table \ref{distance_scales} summarizes
 the scaling parameters which were computed for the measured soft
 X-ray flux but with assumed spin-down conversion efficiencies.

 Distances based on the NE2001 Galactic free electron density model
 are estimated to be, on average, correct within a factor of $\sim
 20\%$. However, as the Galactic free electron distribution is seen to
 vary significantly on small and large scales by more than two orders
 of magnitude it might not be excluded that for single pulsars the
 model can over- or underestimate the free electron density and thus
 yields a wrong distance estimate.  Although it is difficult to
 imagine that the model predicted distance is off by a factor 10 to
 20, a factor of $\sim 3-5$ might not be excluded in rare cases.

 For example, for a pulsar distance of 300pc the NE2001 model predicts
 a dispersion measure $DM=\int_{0}^{300pc} n_e\, dl$ of
 $4.25\,\mbox{pc cm}^{-3}$. If compared with the measured value of
 $34.36\,\mbox{pc cm}^{-3}$ this means that any extra unmodeled
 material along the line of sight to \PSRB\, would need to contribute
 a radio dispersion measure of $\sim 30\,\mbox{pc cm}^{-3}$. This
 correction would not necessarily be due to an unmodeled H-II region
 (typically $DM\sim 1000 \,\mbox{pc cm}^{-3}$) but would be only
 slightly higher than, but still comparable to the typical "clump"
 added in the NE2001 model for discrepant lines of sight, e.g.~towards
 PSR J1031-6117, PSR J1119-6127, and PSR J1128-6219 (Cordes \& Lazio
 2003).  Although these "clumps" in the NE2001 model do not
 necessarily represent physical objects, we did inspect the Digitized
 Sky Survey images and the images from our ESO/NTT observations to see
 whether they show any evidence of enhanced diffuse emission near to
 the region of \PSRB. No noticeable source was recognized that could
 account for this extra correction. Inspecting the Southern H-Alpha
 Sky Survey Atlas (Gaustad et al.~2001) revealed some indications on a
 larger scale of $\sim 10^\circ$ for a H-II region to the north of
 \PSRB, but it is unlikely that this could affect the distance
 estimate by a factor of 5 to 10 or even more (J.Lazio priv.~com.).

 One way to evaluate a possible range of distances is by comparing the
 transverse velocity, derived from the measured proper motion and an
 assumed distance, to typical pulsar velocities. For instance, for the
 1.45 kpc inferred from NE2001, the proper motion measurement of 48.7
 mas yr$^{-1}$ (Hobbs et al.~2004) translates into a transverse speed
 of 335 km s$^{-1}$ which compares to a mean 2-D speed as found for a
 large sample of pulsars of $246\pm22$ km s$^{-1~}$ (Hobbs et
 al.~2005). For the larger distance of 2.15 kpc derived using the
 Taylor \& Cordes (1993) model, the velocity increases to a rather
 large but still plausible value of 497 km s$^{-1}$. If the pulsar
 were a factor of 5 or 22 closer, than the transverse speed drops to
 67 km s$^{-1}$ and 15 km s$^{-1}$, respectively. Unless the
 non-measurable radial speed happens to be large, the latter 
 velocity is extremely small and quite unlikely.

 Although the distance of the pulsar might be the most obvious
 parameter which may lead to a too high X-ray conversion efficiency,
 there might be others. The conversion efficiency here and in Becker
 \& Tr\"umper (1997) is for isotropy.  Would it be possible that the
 pulsar has an atypically small beaming angle?  or that it could be a
 coaction of several effects like beaming, distance, peculiarities of
 its magnetosphere which cause the apparent higher X-ray efficiency?

 So far \PSRB\, was not included in any radio parallax measurement
 program.  The pulsar's radio emission is strong enough so that a
 radio parallax could be measured with the VLBA. This will be done on
 a time scale of about two years and will provide us with the missing
 information on whether it is "simply" the pulsar distance which
 causes the inferred huge X-ray conversion or whether \PSRB\, bears
 more interesting things to discover.

\acknowledgments

WB acknowledges that a significant part of the XMM-data analysis of
PSR B0628-28 was done during the workshop on {\em High Energy
Phenomena of Compact Objects} which was held from March $7-18$ 2005 at
the Theoretical Institute for Advanced Research in Astrophysics
(TIARA) in Hsinchu/Taiwan. WB further acknowledges discussion with
R.Mignani, J.M.~Cordes and T.J.W.~Lazio. We are grateful to Olaf Maron
for the use of his database of pulsar radio fluxes.
Optical observations were made with ESO Telescopes at the La Silla 
Observatories under program 64.N-0698. We thank the anonymous referee 
for thoroughly reading the manuscript and the many useful comments.

\clearpage

\begin{deluxetable}{lc}
\tablewidth{0pc}
\tablecaption{Ephemerides of \PSRB \label{t:radio}}
\tablehead{}
\startdata

Right Ascension (J2000)                                              & $06^h 30^m 49^s\!.417 \pm 00^s\!.006$  \\
Declination (J2000)                                                  & $-28^d 34^m 42^s\!.91 \pm 00^s\!.1$    \\
First date for valid parameters (MJD)                                & 52956                            \\
Last date for valid parameters (MJD)                                 & 53256                            \\
Infinite-frequency geocentric pulse arrival time (MJD, UTC)$^a$      & 53106.000013822                  \\
Pulsar rotation period ($s$)                                         & 1.24442260399                    \\
Pulsar rotation frequency ($s^{-1}$)                                 & 0.8035855317903                  \\
First derivative of pulsar frequency ($s^{-2}$)                      & $-4.61639 \times 10^{-15}$       \\
Second derivative of pulsar frequency ($s^{-3}$)                     & $5.30 \times 10^{-29}$           \\
Spin-down age (yr/$10^6$)                                            & 2.754                            \\
Spin-down energy ($\mbox{erg/s}/10^{32}$)                            & 1.445                            \\
Inferred Magnetic Field ($G/10^{12}$)                                & 3.02                             \\
Dispersion Measure ($\mbox{pc/cm}^3$)                                & 34.36                            \\
Distance$^b$ (kpc)                                                   & 1.45                             \\
\enddata
\tablecomments{\newline
$^a$ The integer part of this time is the barycentric (TDB) epoch of
 RA, DEC, f, $\dot{f}, \ddot{f}$.\newline
$^b$ Dispersion-measure inferred distance according to Cordes and Lazio (2002).}
\end{deluxetable}

\clearpage

\begin{deluxetable}{llccccc}
\tablewidth{0pc}
\tablecaption{Instrument setups, filter usage, start time, durations, and effective exposures of
the XMM-Newton observations of \PSRB.\label{t:xray_obs}}
\tablehead{}
\startdata
Detector & \quad\quad\quad Mode & Filter & Start time & Duration & eff.~Exp. \\
{} & {} & {} & (UTC) & (s) & (s) \\\hline\\[-1ex]

EMOS1 & PrimeFullWindow  & Medium & 2004-02-28T02:20:13  & 48\,463 & 35\,568 \\
EMOS2 & PrimeFullWindow  & Medium & 2004-02-28T02:20:10  & 48\,471 & 34\,590 \\
EPN   & PrimeLargeWindow & Thin   & 2004-02-28T02:39:43  & 46\,998 & 31\,738 \\
\enddata
\end{deluxetable}

\clearpage

\begin{deluxetable}{ccccccc}
\tablewidth{0pc}
\tablecaption{Models and parameters as fitted to the XMM-Newton observed X-ray spectrum of \PSRB
\label{spectral_fits_J0628}}
\tablehead{}
\startdata
 model$^a$ & $\chi_\nu^2$ & $\nu$   & $N_H/10^{20}$       &          $\alpha$ / $kT$$^b$         &      Normalization at 1 keV         &  Radius$^c$ \\
  {}    &      {}         &   {}    & $\mbox{cm}^{-2}$    &                     {}               &      Photons/keV/cm${^2}$/s         &      m      \\\hline\\[-1ex]

  PL    &      0.926     &   37    & $6.0_{-1.8}^{+3.0}$ &   $2.63_{-0.15}^{+0.22}$             & $1.0_{-0.1}^{+0.13}\times 10^{-5}$  &      {}                      \\\\[-1.5ex]

  BB    &      1.749     &   37    & $\le {0.2}$        &   $0.23_{-0.01}^{+0.01}$             &               {}                    &   $131.7_{-12.5}^{ +13.8}$   \\\\[-1.5ex]

 BB+BB  &      0.808     &   35    & $\le {0.9}$        &\parbox{3cm}{$\mbox{kT}_1=0.142_{-0.01}^{+0.22}$\\[1ex]
                                                            $\mbox{kT}_2=0.402_{-0.04}^{+0.09}$ } &        {}                    & \parbox{1.8cm}{$278.8_{-59.2}^{ +116.4}$\\[1ex]
                                                                                                                                    $34.8_{-11.33}^{+9.75}$} \\\\[-1.5ex]

 BB+PL  &      0.810     &   35   & $\le {2.5}$         & $2.27^{+0.23}_{-0.13}\, /\, 0.25_{-0.04}^{+0.05}$& $4.8^{+1.6}_{-0.9} \times10^{-6}$ &     $69_{-25}^{+30}$  \\\\[-1.5ex]

 BB+PL  &      0.969     &   35   & $6.4_{-1.4}^{+4.6}$  & $2.26_{-0.2}^{+0.3}\, /\, < 0.053$    & $1.0_{-0.1}^{+0.2}\times 10^{-5}$ &             $10\,000$             \\\\[-1.5ex]

 BKNPL  &      0.935     &   37   &  $\le {1.4}$        & \parbox{3cm}{$\alpha_1=1.478_{-0.02}^{+0.02}$\\[1ex]
                                                            $\alpha_2=2.7215_{-0.3}^{+0.4}$}                     & $1.2_{-0.1}^{+0.2}\times 10^{-5}$ &        {}   \\\\[-1ex] \hline

\enddata
\tablecomments{
$^a$ BB = blackbody; PL = power law;  BKNPL = broken power law. Errors represent the $1\sigma$ confidence
range. For single component spectral models errors were computed for one parameter of interest. For 
composite models the errors were computed for two parameters of interest.\\
$^b$ The entry in this column depends on the spectral model --- it is the
power law photon index $\alpha$ or the temperature $kT$ in keV\\
$^c$ For thermal models for which we computed or fixed the radius of the emitting area we
assumed a pulsar distance of 1.45 kpc.}
\end{deluxetable}

\clearpage

\begin{deluxetable}{rcrr}
\tablewidth{0pc}
\tablecaption{Distance scales of \PSRB\, for assumed spin-down\newline
luminosity to X-ray conversion factors $L_x/\dot{E}$
\label{distance_scales}}
\tablehead{}
\startdata
   $L_x/\dot{E}\,\,$ &    $f_x(0.1-2.4$ keV)        & $d\,\,\,\,$    &  $d_{DM} / d$ \\
         {}          &   $\mbox{erg/s/cm}^2$        &  pc\,\,        &    {}      \\\hline\\[-1ex]

   $5\times 10^{-3}$ &   $7.1394 \times 10^{-14}$    &  290.8        &     4.98  \\
   $5\times 10^{-3}$ &   $9.4194 \times 10^{-14}$    &  253.3        &     5.72   \\
   $5\times 10^{-3}$ &   $14.412 \times 10^{-14}$    &  204.7        &     7.07   \\[1ex]

   $        10^{-3}$  &     $7.1394 \times 10^{-14}$  &  130.1       &    11.13  \\
   $        10^{-3}$  &     $9.4194 \times 10^{-14}$  &  113.3       &    12.78   \\
   $        10^{-3}$  &     $14.412 \times 10^{-14}$  &   91.5       &    15.82   \\[1ex]

   $5\times 10^{-4}$  &     $7.1394 \times 10^{-14}$  &   92.9       &    15.58  \\
   $5\times 10^{-4}$  &     $9.4194 \times 10^{-14}$  &   80.1       &    18.08   \\
   $5\times 10^{-4}$  &     $14.412 \times 10^{-14}$  &   64.7       &    22.38   \\[0.5ex]\hline
\enddata
\end{deluxetable}

\clearpage

\clearpage

\clearpage

\clearpage

\clearpage

\clearpage

\clearpage

\clearpage

\clearpage

\end{document}